\documentclass[12pt]{iopart}

\usepackage{graphicx}
\usepackage{cite}
\usepackage{hyperref}
\usepackage{bm}

\begin{document}

\title[]{Spin and polarization effects on the nonlinear Breit-Wheeler pair production in laser-plasma interaction}

\author{Huai-Hang Song$^{1,6}$, Wei-Min Wang$^{2,1,8}$, Yan-Fei Li$^3$, Bing-Jun Li$^3$, Yu-Tong Li$^{1,6,7,9}$, Zheng-Ming Sheng$^{4,5,8,10}$, Li-Ming Chen$^{4,8,10}$, and Jie Zhang$^{4,1,8}$}

\address{$^1$ Beijing National Laboratory for Condensed Matter Physics, Institute of Physics, Chinese Academy of Sciences, Beijing 100190, China}
\address{$^2$ Department of Physics and Beijing Key Laboratory of Opto-electronic Functional Materials and Micro-nano Devices, Renmin University of China, Beijing 100872, China}
\address{$^3$ Department of Nuclear Science and Technology, Xi’an Jiaotong University, Xi’an 710049, China}
\address{$^4$ Key Laboratory for Laser Plasmas (MoE) and School of Physics and Astronomy, Shanghai Jiao Tong University, Shanghai 200240, China}
\address{$^5$ SUPA, Department of Physics, University of Strathclyde, Glasgow G4 0NG, United Kingdom}
\address{$^6$ School of Physical Sciences, University of Chinese Academy of Sciences, Beijing 100049, China}
\address{$^7$ CAS Center for Excellence in Ultra-intense Laser Science, Shanghai 201800, China}
\address{$^8$ Collaborative Innovation Center of IFSA, Shanghai Jiao Tong University, Shanghai 200240, China}
\address{$^9$ Songshan Lake Materials Laboratory, Dongguan, Guangdong 523808, China}
\address{$^{10}$ Tsung-Dao Lee Institute, Shanghai Jiao Tong University, Shanghai 200240, China}

\ead{weiminwang1@ruc.edu.cn and ytli@iphy.ac.cn} \vspace{10pt}

\begin{abstract}
The spin effect of electrons/positrons ($e^-$/$e^+$) and polarization effect of $\gamma$ photons are investigated in the interaction of two counter-propagating linearly polarized 10-PW-class laser pulses with a thin foil target. The processes of nonlinear Compton scattering and nonlinear Breit-Wheeler pair production based on spin- and polarization-resolved probabilities are implemented into the particle-in-cell (PIC) algorithm by Monte Carlo methods. It is found from PIC simulations that the average degree of linear polarization of emitted $\gamma$ photons can exceed $50\%$. This polarization effect leads to reduced positron yield by about $10\%$. At some medium positron energies, the reduction can reach $20\%$. Furthermore, we also observe that the local spin polarization of $e^-$/$e^+$ leads to a slight decrease of the positron yield about $2\%$ and some anomalous phenomena about the positron spectrum and photon polarization at the high-energy range, due to spin-dependent photon emissions. Our results indicate that spin and polarization effects should be considered in calculating the pair production and laser-plasma interaction with the laser power of 10-PW class.
\end{abstract}

%
%
%
%
%

\section{Introduction}

Over the past decades, the intensity of lasers has increased rapidly \cite{Cartlidge2018science,Danson2019hplse} with the laser technical progress based on chirped pulse amplification \cite{Strickland1985oc}. Several multi-petawatt (PW) \cite{Sung2017oe} and 10-PW-class \cite{Tanaka2020mre,Lureau2020hplse} femtosecond laser systems have been built, which are expected to achieve an unprecedented peak power density up to the order of $10^{23-24}$ W/cm$^{2}$ with tightly focusing. Such ultraintense lasers enable laser plasma interactions to enter the quantum electrodynamics (QED) regime \cite{Piazza2012rmp,Mourou2019rmp,Blackburn2020rmpp}. Electrons experiencing the ultra-intense transverse field can stochastically radiate $\gamma$ photons by nonlinear Compton scattering and lose a considerable amount of energy if the quantum parameter $\chi_e=(e\hbar/m_e^3c^4)|F_{\mu\nu}p^{\nu}|\sim1$ \cite{Ritus1985jslr,Berestetskii1982qed,Baier1998qed}, where $F_{\mu\nu}$ is the field tensor, $p^{\nu}$ is the electron four-momentum, and the constants $\hbar$, $m_e$, $e$ and $c$ are the reduced Planck constant, the electron mass and charge, and the speed of light, respectively. As another cross channel of the same reaction, $\gamma$ photons traveling through the ultraintense field possibly further decay into electron-positron ($e^-e^+$) pairs by nonlinear Breit-Wheeler process \cite{Breit1934pr} with another characteristic parameter $\chi_\gamma=(e\hbar^2/m_e^3c^4)|F_{\mu\nu}k^{\nu}|$, where $\hbar k^{\nu}$ is the photon four-momentum. This sort of pair production by light-by-light scattering was first demonstrated by the famous SLAC E-144 experiment \cite{Burke1997prl} in 1990s, where only  about one hundred positrons was detected due to the limitation of the laser intensity at that time. The laser pulse and electron beam collisions utilizing today's high-intensity laser facilities in all-optical setups are also studied recently \cite{Sokolov2010prl,Blackburn2017pra}.

	With upcoming 10-PW-class lasers \cite{Lureau2020hplse}, abundant $e^-e^+$ pairs can even be produced in laser-plasma interactions without the need to pre-accelerate electrons to GeV energies. When such an ultraintense laser irradiates plasmas, electrons would be accelerated to ultrarelativistic energies and deflected by laser or strong self-generated fields in plasmas to gain a Lorentz boosted field strength in the electron's moving frame to achieve $\chi_e\sim1$. Many theoretical proposals for producing dense $e^-e^+$ pairs or even avalanche-like cascades have been put forward,
such as through the laser collision configuration seeded by electrons/positrons ($e^-$/$e^+$) \cite{Bell2008prl,Nerush2011prl,Grismayer2017pre} or plasmas \cite{Zhu2016nc}, and directly laser-solid interactions \cite{Ridgers2012prl,Kostyukov2016pop, Sorbo2018njp,Wang2017pre}.
Generating copious positrons or dense $e^-e^+$ plasmas in the laboratory is of great importance in astrophysics \cite{Goldreich1969apj,Piran2005rmp,Lobet2015prl}, nuclear physics \cite{Ruffini2010pr}, and materials science \cite{Danielson2015rmp}.

Moreover, the $e^-e^+$ spin effect and $\gamma$-photon polarization effect have aroused interest in the strong-field QED regime \cite{King2013pra,Buscher2020hplse,Seipt2020pra}.
An ultrarelativistic electron beam is found to be transversely spin-polarized by a single-shot collision of an elliptically polarized laser pulse \cite{Li2019prl} or a two-color laser pulse \cite{Song2019pra,Seipt2019pra} due to hard $\gamma$ photon emissions, analogous to the Sokolov-Ternov effect \cite{Sokolov1968qed,Baier1967pla} in the magnetic field. Similar spin polarization processes for newly created $e^-e^+$ pairs are also investigated \cite{Wan2020plb,Chen2019prl,Li2020prl2}. The general view is that constructing an asymmetric laser field is the key to realize spin-polarized electrons or positrons. The emitted $\gamma$ photons could be polarized via nonlinear Compton
scattering \cite{King2016pra,Li2020prl1,Xue2020mre}, whose polarization strongly depends on the initial spin of electrons \cite{Li2020prl1}. It is found that only linearly polarized $\gamma$ photons can be generated by unpolarized or transversely polarized electrons \cite{King2013pra,Li2020prl1}. 

A more sophisticated description for $e^-e^+$ pair production by taking into account the $e^-e^+$ spin and $\gamma$-photon polarization has been discussed \cite{King2013pra,Wan2020prr,Seipt2020arxiv},  based on single-particle model analyses or simulations. The photon polarization
is shown to significantly reduce the pair yield by a factor of over $10\%$ in the collision of ultraintense laser pulse and electron beam \cite{Wan2020prr}. In the rotating electric fields, the growth rate of the $e^-e^+$ cascade is also found to be suppressed \cite{Seipt2020arxiv}. However, it still demands to be studied that in the laser-plasma interaction to what extent the $e^-e^+$ spin and $\gamma$-photon polarization impact on the positron yield.

In this paper, we study the $e^-e^+$ pair production in the laser-plasma interaction, by taking into account the spin of $e^-/e^+$ and the polarization of $\gamma$ photons. The spin- and polarization-resolved probabilities of nonlinear Compton scattering and nonlinear Breit-Wheeler pair production are both implemented into the widely employed QED particle-in-cell (PIC) algorithm, which can self-consistently capture the collective plasma dynamics and QED related processes. Here, we focus on the pair production in the interaction of two counter-propagating linearly polarized laser pulses of the same frequency and intensity with a thin foil target. This is a particularly advantageous configuration under the current laser intensity for triggering the QED pair production,  due to the formation of the linearly-polarized electromagnetic standing wave (EMSW) \cite{Jirka2016pre,Grismayer2017pre}. Our simulation results show that the positron yield is reduced by about $10\%$ with the spin and polarization effects included. This significant difference is primarily caused by the polarized intermediate $\gamma$ photons with an average linear-polarization degree of more than $50\%$. In addition, we also observe a decrease of positron number by about $2\%$ and some anomalous phenomena for high-energy particles due to local spin polarization of $e^-/e^+$. This work indicates that the previously widely adopted spin- and polarization-averaged probabilities implemented in QED-PIC codes cannot accurately calculate the positron yield in the laser-plasma interaction for 10-PW-class lasers, and spin and polarization effects should be considered.

\section{Theoretical model}
\label{theory}

In order to determine the spin of electron after the photon emission and also the polarization of the emitted $\gamma$ photon, the spin- and polarization-resolved photon emission probability is employed, which is derived in the Baier-Katkov QED operator method \cite{Baier1998qed,Li2020prl1},
\begin{eqnarray}\label{eq1}
\frac{d^2W_{rad}}{dudt}&=&\frac{\alpha m^2c^4}{4\sqrt{3}\pi \hbar \varepsilon_e}\left\{\frac{u^2-2u+2}{1-u}K_{\frac{2}{3}}(y)-{\rm Int}K_{\frac{1}{3}}(y) -uK_{\frac{1}{3}}(y)(\bm S_i \cdot \bm e_2) \right.\nonumber\\
&&\left. + \left[ 2K_{\frac{2}{3}}(y)-{\rm Int}K_{\frac{1}{3}}(y)\right](\bm S_i \cdot \bm S_f)-\frac{u}{1-u}K_{\frac{1}{3}}(y)(\bm S_f \cdot \bm e_2)\right.\nonumber\\
&&\left. +\frac{u^2}{1-u}\left[K_{\frac{2}{3}}(y)-{\rm Int}K_{\frac{1}{3}}(y)\right](\bm S_i \cdot \bm e_v)(\bm S_f \cdot \bm e_v) \right.\nonumber\\
&&\left. +\frac{u}{1-u} K_{\frac{1}{3}}(y)(\bm S_i \cdot \bm e_1) \xi_1+ \left[ \frac{2u-u^2}{1-u} K_{\frac{2}{3}}(y) -u{\rm Int}K_{\frac{1}{3}}(y) \right] (\bm S_i \cdot \bm e_v)\xi_2 \right.\nonumber\\
&&\left. + \left[ K_{\frac{2}{3}}(y) - \frac{u}{1-u} K_{\frac{1}{3}}(y)(\bm S_i \cdot \bm e_2) \right]\xi_3 \right\},
\end{eqnarray}
where $K_{\nu}(y)$ is the second-kind modified Bessel function of the order of $\nu$, ${\rm Int}K_{\frac{1}{3}}(y) \equiv \int_{y}^{\infty}K_{1/3}(x)dx$, $y=2u /[3(1-u)\chi_e]$, $u=\varepsilon_\gamma / \varepsilon_e$, $\varepsilon_e$ the electron energy before the photon emission, $\varepsilon_\gamma$ the emitted photon energy, and $\alpha\approx 1/137$ the
fine structure constant. $\bm e_v$ is the unit vector along the electron velocity, $\bm e_1$ is the unit vector along the electron transverse acceleration, and $\bm e_2=\bm e_v \times \bm e_1$. $\bm S_i$ and $\bm S_f$ are the spin vectors
of an electron before and after photon emission, respectively, with $|\bm S_{i,f}|=1$. The photon polarization is represented by Stokes parameters $\bm \xi=(\xi_1, \xi_2, \xi_3)$, defined with respect to the basis vector ($\bm e_1,\bm e_2,\bm e_v$). The case $\xi_1=\xi_2=0$, $\xi_3=1(-1)$ means the photon is linearly polarized along $\bm e_1(\bm e_2)$, and the case $\xi_1=\xi_3=0$, $\xi_2=\pm 1$ means the photon is circularly polarized. After averaging over $\bm S_i$ and summing up over $\bm S_f$ and $\bm \xi$, the widely employed spin- and polarization-averaged photon emission probability is obtained \cite{Nerush2011prl,Ridgers2014jcp}. Note that the above description can also be applied to the positron.

For convenience, we define high-energy photons with $u/(1-u)\ge1$ and low-energy photons with $u/(1-u)\ll1$. We ignore the correlation terms involving both $\bm S_f$ and $\bm \xi$ in equation~(\ref{eq1}), because the final electron spin and photon polarization are calculated separately in our Monte-Carlo method \cite{Cain}.

Utilizing the similar method, the probability of the pair production with the photon polarization  included can be written as \cite{Baier1998qed,Wan2020prr,Li2020prl2},
\begin{eqnarray}\label{eq2}
\frac{d^2W_{pairs}}{d\varepsilon_{+} dt}&=&\frac{\alpha m^2c^4}{\sqrt{3}\pi \hbar \varepsilon_\gamma^2}\left\{\frac{\varepsilon_{+}^2+\varepsilon_{-}^2}{\varepsilon_{+}\varepsilon_{-}}K_{\frac{2}{3}}(y)+{\rm Int}K_{\frac{1}{3}}(y) -\xi_3'K_{\frac{2}{3}}(y) \right\},
\end{eqnarray}
where $y=2\varepsilon_\gamma^2/(3\chi_\gamma\varepsilon_{+}\varepsilon_{-})$, $\varepsilon_-$ and $\varepsilon_+$ are the energies of the produced electron and positron, respectively. The last term containing $\xi_3'$ in equation~(\ref{eq2}) accounts for the photon polarization effect on the pair production. It is apparent that if $\xi_3'$ is a positive value, the pair production probability $d^2W_{pairs}/(d\varepsilon_{+} dt)$ is reduced and consequently the positron yield decreases. The decrease can be up to $30\%$ for medium-energy positrons \cite{Wan2020prr}. The Stokes parameters need to be transformed from the photon emission frame ($\bm e_1,\bm e_2,\bm e_v$) to the pair production frame ($\bm e_1',\bm e_2',\bm e_v$) to obtain  a new set of Stokes parameters $(\xi_1', \xi_2', \xi_3')$ that required in equation~(\ref{eq2}), through the matrix rotation \cite{McMaster1961rmp}
\begin{eqnarray}\label{eq3}
\xi_1' &=& \xi_1\cos(2\theta)-\xi_3\sin(2\theta),\nonumber\\
\xi_2' &=& \xi_2,\\
\xi_3' &=& \xi_1\sin(2\theta)+\xi_3\cos(2\theta),\nonumber
\end{eqnarray}
where $\bm e_1'$ is the unit vector along ${\bm E}+\bm e_v \times \bm B-\bm e_v \cdot (\bm e_v \cdot \bm E)$, $\bm e_2'=\bm e_1' \times \bm e_v$, and $\theta$ is the angle between $\bm e_1$ and $\bm e_1'$.

The semiclassical formulas of photon emission probability in equation~(\ref{eq1}) and pair production probability in equation~(\ref{eq2}) are derived based on the local constant field approximation \cite{Piazza2018pra,Ilderton2019pra}, which is justified at an ultraintense laser intensity of $a_0=|e|E_L/mc\omega_0\gg1$, where $\omega_0$ is the laser frequency. The stochastic photon emission by an electron or a positron and pair production by a $\gamma$ photon are calculated using the standard QED Monte-Carlo algorithm \cite{Elkina2011prab,Ridgers2014jcp,Green2015cpc,Gonoskov2015pre,Wang2016arxiv} but with the spin- and polarization-resolved probabilities. The $e^-e^+$ dynamics in the external electromagnetic field is described by classical Newton-Lorentz equations, and their spin dynamics are calculated according to the Thomas-Bargmann-Michel-Telegdi equation \cite{Bargmann1959prl,Baier1972spu}. Detailed Monte-Carlo methods for numerical modeling of spin and polarization we employ can be found in Refs.~\cite{Li2019prl,Li2020prl1,Cain,Li2020prl2}. In our simulations, spin vectors of newly created pairs are also included, which however has a relatively weak influence on the pair production in the linearly-polarized EMSW.

\section{Simulation results and analysis}
\subsection{Simulation setup}

We implement the spin- and polarization-resolved probabilities in equations~(\ref{eq1}) and (\ref{eq2}) into the two-dimensional QED-PIC code by the Monte-Carlo method as described in section~\ref{theory}, to self-consistently study the $e^-e^+$ spin and $\gamma$ photon polarization effects on the pair production in the laser-plasma interaction. The standard QED modules have been benchmarked following Refs.~\cite{Wang2016arxiv,Wang2017pre}, and benchmarks about the spin and polarization modules are presented in \ref{appendix}. In the following simulations, we can artificially switch on or off these two modules for better identifying their impacts by comparison. In our simulation setups, two counter-propagating laser pulses with the same profile of $a_L=a_0\,{\rm sin}^2(\pi t/\tau_0)\times{\rm exp}{(-r^2/r^2_0)}$ within $0< t \leq\tau_0$ are normally incident from the left and right boundaries, respectively, and they are both linearly polarized along the $y$ axis. We take the laser normalized peak strength $a_0=800$ (peak intensity $I_0\approx8.9\times 10^{23}$ W/cm$^2$), spot size $r_0=2\lambda_0$, and pulse duration $\tau_0=10T_0$, where $\lambda_0=1\mu$m is the laser wavelength and $T_0=2\pi/\omega_0\approx3.33$ fs is the laser period. A $1\mu $m-thickness fully ionized foil target, composed of electrons and carbon ions, is initially placed in the laser overlapping center of $3.5\lambda_0<x<4.5\lambda_0$ with an electron density of $n_e=50n_c$, where $n_c=m_e\omega_0^2/4\pi e^2$ is the critical density. The computational domain has a size of $8\lambda_0\times 15\lambda_0$ in $x\times y$ directions with $384\times 720$ cells. Each cell contains 100 macro electrons and 16 macro carbon ions. Absorbing boundary conditions are used for both particles and fields in any direction.

In the first simulation case, labeled as {\bf case (\romannumeral1)}, the spin and polarization effects are both incorporated. As reference cases, we have also performed another three simulations under the same physical parameters as those in {\bf case (\romannumeral1)} except that: in {\bf case (\romannumeral2)}, the spin and polarization effects are not included, which is the widely adopted method in the current QED-PIC codes \cite{Elkina2011prab,Ridgers2014jcp,Gonoskov2015pre}; in {\bf case (\romannumeral3)}, only spin effect is included, where the polarization-dependent terms are summed up over in equation~(\ref{eq1}) and averaged over in equation~(\ref{eq2}), while retaining spin-dependent terms in these two probability equations; in {\bf case (\romannumeral4)}, we switch off the photon annihilation and pair production processes but still include the spin and polarization effects, to check the original polarization characteristics of emitted $\gamma$ photons.


\begin{figure*}[t]
\centering
\includegraphics[width=0.9\textwidth]{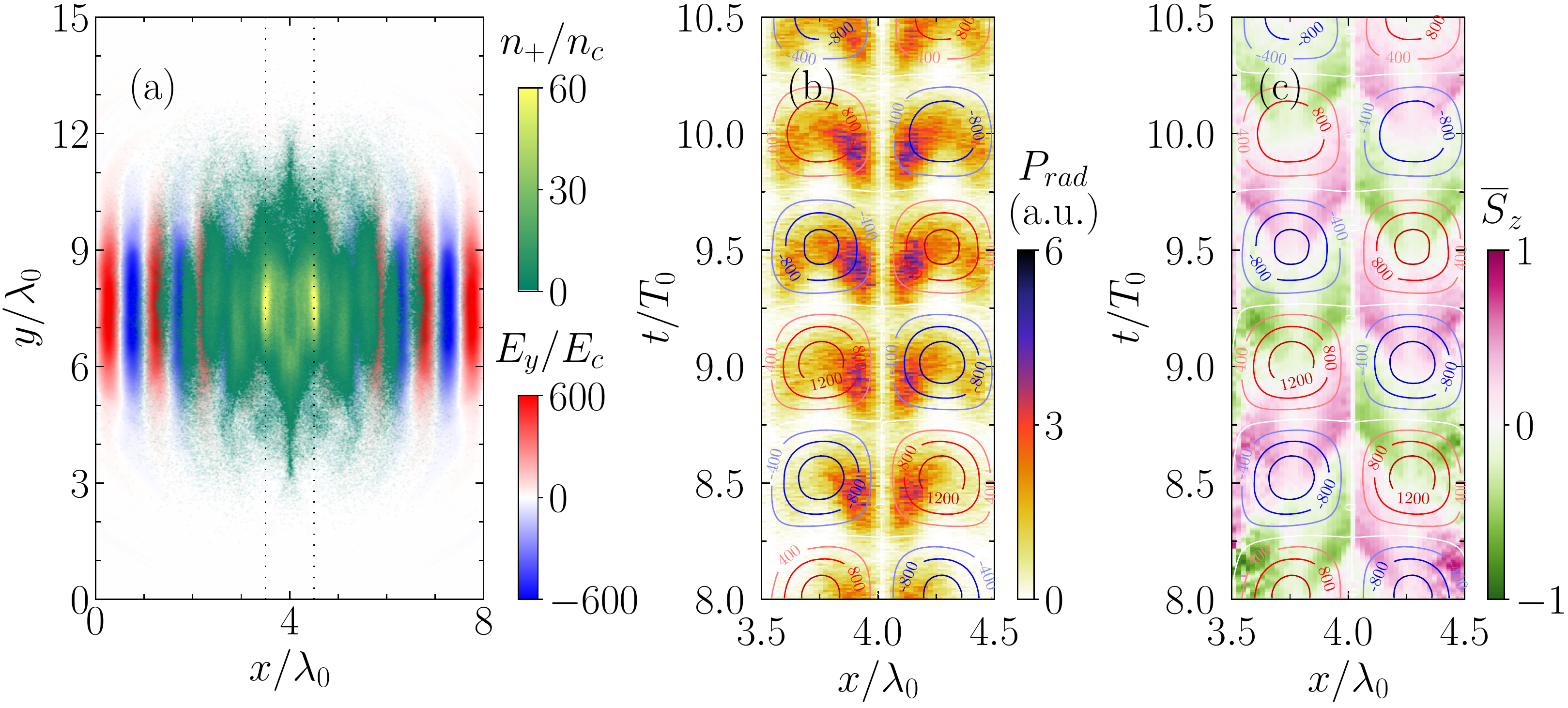}
\caption{\label{fig1} Simulation results of {\bf case (\romannumeral1)}. (a) Spatial distribution of the density of produced positrons $n_+$, together with the electric field $E_y$ of two counter-propagating laser pulses at $t=12T_0$, where $E_c$ is $mc\omega_0/|e|$. The dotted lines outline the initial boundaries of the foil target. Space-time evolution of (b) the radiation power $P_{rad}$ of photons with energies above 100 MeV and (c) the average electron spin degree $\overline S_z$ along $y=7.5\lambda_0$. The laser magnetic field $B_z$ normalized by $mc\omega_0/|e|$ is also shown by contour lines to outline the magnetic region of the EMSW both in (b) and (c).}
\end{figure*}

\subsection{Simulation results}
For {\bf case (\romannumeral1)}, figure~\ref{fig1}(a) illustrates the spatial distribution of the positron density $n_{+}$ at the end of the laser-foil interaction at $t=12T_0$ when two laser pulses have passed through each other. We can see that dense and spatially modulated \cite{Esirkepov2015pla} positrons are produced, with a maximum density of $n_+=60n_c$, already exceeding the initial electron density of the foil target. A transient linearly EMSW responsible for the abundant positron production is constructed by two counter-propagating linearly polarized laser pulses in the time interval $7T_0<t<11T_0$, covering the entire foil plasma zone. The space-time evolution of magnetic field component $B_z$ of EMSW along $y=7.5\lambda_0$ is shown by contour lines both in figures~\ref{fig1}(b) and \ref{fig1}(c). The formed EMSW is divided into electric region (maximize at magnetic nodes $x=m\lambda_0/2$ and $t=n T_0/2+T_0/4$) and magnetic region (maximize at magnetic antinodes $x=m\lambda_0/2+\lambda_0/4$ and $t=n T_0/2$), where $m$ and $n$ are integers. These two distinct regions are shifted by $\lambda_0/4$ spatially and $T_0/4$ temporally ($\pi/2$ phase offset). Furthermore, electrons are accelerated or decelerated by the electric field along the $y$ direction in the electric region, while emitting photons and simultaneously losing energies primarily in the magnetic region \cite{Kostyukov2016pop}. The photon radiation power $P_{rad}$ shown in figure~\ref{fig1}(b) verifies that more high-energy photons are emitted in the early stage after entering the magnetic region. The previous study \cite{Jirka2016pre} has presented that this type of field configuration is favorable for improving quantum parameter $\chi_e$ with 10-PW-class laser facilities, and consequently are the photon emission and pair production. The quantum parameter $\chi_e$ can reach a maximum of 3 in our case. At the end of the simulation, about $30\%$ laser energies are absorbed, among which, $24\%$ are transformed into photons, $5.5\%$ into electrons and positrons, and less than $0.5\%$ into ions.

\begin{figure*}[t]
\centering
\includegraphics[width=6.3in]{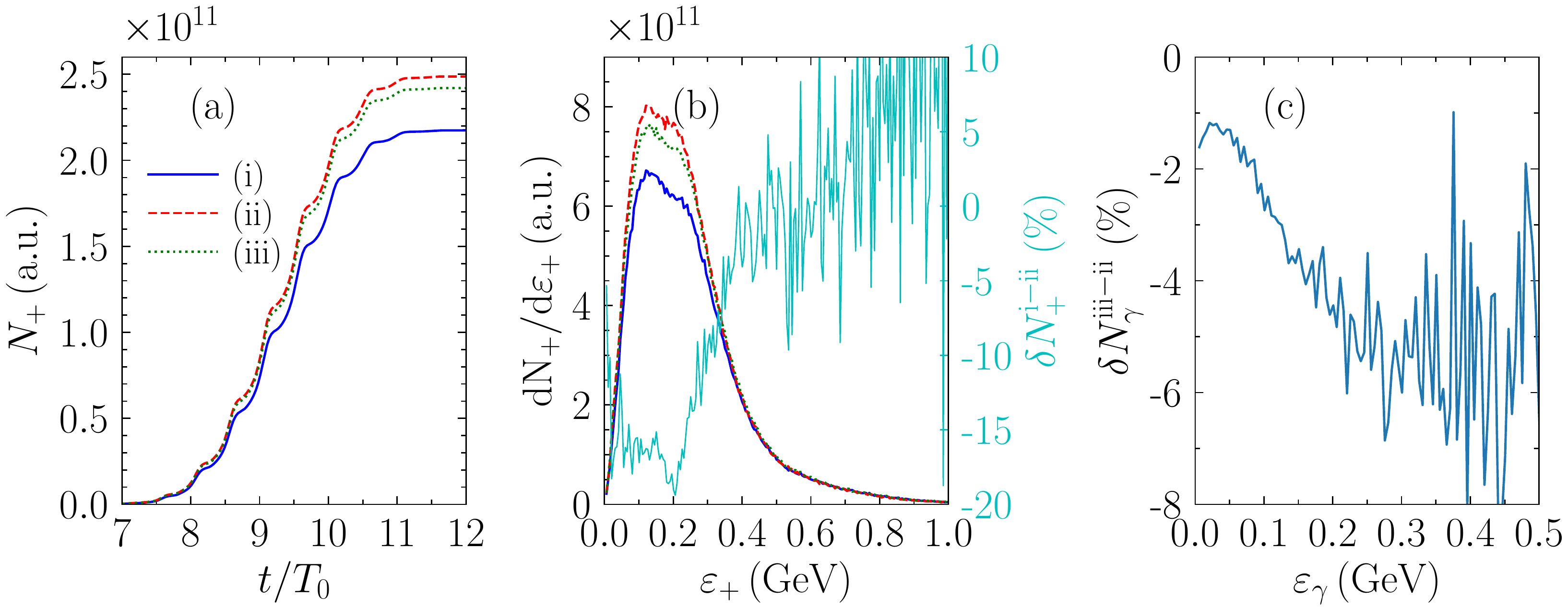}
\caption{\label{fig2} (a) The time evolution of total positron number $N_+$ in three simulation cases: {\bf case (\romannumeral1)} with spin and polarization effects (blue solid line), {\bf case (\romannumeral2)} without spin and polarization effects (red dashed line), and {\bf case (\romannumeral3)} with only spin effect (green dotted line). (b) Positron energy spectrum $dN_+/d\varepsilon_+$ under the three cases at the end of the laser interaction. The relative deviation $\delta N_+^{\rm \romannumeral1-\romannumeral2}=(dN_+^{\rm\romannumeral1}/d\varepsilon_+-dN_+^{\rm\romannumeral2}d\varepsilon_+)/dN_+^{\rm\romannumeral2}d\varepsilon_+$ between {\bf case (\romannumeral1)} and {\bf case (\romannumeral2)} is also presented by the cyan line. (c) The relative deviation $\delta N_\gamma^{\rm \romannumeral3-\romannumeral2}=(dN_\gamma^{\rm\romannumeral3}/d\varepsilon_\gamma-dN_\gamma^{\rm\romannumeral2}/d\varepsilon_\gamma)/dN_\gamma^{\rm\romannumeral2}d\varepsilon_\gamma$ of photon number versus photon energy $\varepsilon_\gamma$ between {\bf case (\romannumeral3)} and {\bf case (\romannumeral2)}.}
\end{figure*}

The time evolution of the total positron number $N_+$ of cases (\romannumeral1)-(\romannumeral3) is illustrated in figure~\ref{fig2}(a). During the existence of EMSW in the time interval of $7T_0<t<11T_0$, $N_+$ increases dramatically. The stair-step-like growth with a period of $0.5T_0$ is attributed to the fact that electrons strongly emit photons mostly in magnetic regions as already shown in figure~\ref{fig1}(b). The most important feature is that when the spin and polarization effects are fully considered in {\bf case (\romannumeral1)}, the total number of positrons is reduced by $12\%$ compared with the {\bf case (\romannumeral2)} that excluding the two effects, i.e. $\Delta N_{+}^{\rm\romannumeral1-\romannumeral2}=(N_+^{\rm\romannumeral1}-N_+^{\rm\romannumeral2})/N_+^{\rm\romannumeral2}\approx-12\%$. Furthermore, the relative deviation of positron number $\delta_+^{\rm \romannumeral1-\romannumeral2}$ also depends on the positron energy $\varepsilon_+$ shown in figure~\ref{fig2}(b), exhibiting a maximum difference as large as $-20\%$ at $\varepsilon_+=200$ MeV. The difference of positron yield is mainly attributed to the linear polarization of emitted $\gamma$ photons, which will be detailed in the next subsection.

Then, we analyze the spin dynamics of electrons in the linearly-polarized EMSW. As emitting a high-energy photon, the electron spin more probably flips to the direction antiparallel to the magnetic field in the electron’s rest frame, according to equation~(\ref{eq1}) [see figure~4(b) in reference~\cite{Song2019pra}]. In the magnetic region where photon emissions are concentrated on, the spin-flip trend of the electron is determined by the direction of magnetic field $B_z$. More specifically, the electron spin is more likely antiparallel to the $B_z$ direction after the photon emission (and positron spin is more likely parallel to that). This is evidenced by the time-space evolution of the average spin component $\overline S_z$ of electrons shown in figure~\ref{fig1}(c). It shows that $\overline S_z$ is temporally oscillating with a laser frequency $\omega_0$, but the total degree of spin still remains nearly zero due to the symmetry of the field, which is a kind of local spin polarization.

\begin{figure*}[t]
\centering
\includegraphics[width=0.9\textwidth]{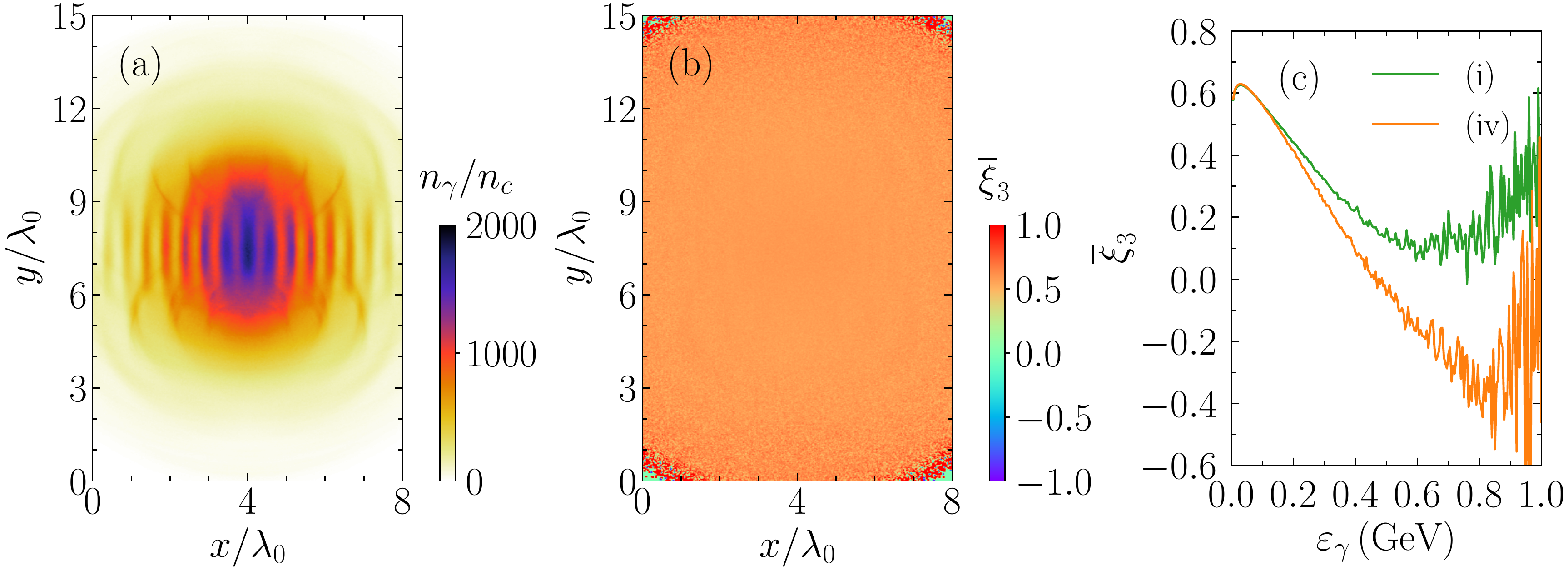}
\caption{\label{fig3} Spatial distribution of (a) photon density $n_\gamma$ and (b) average linear-polarization degree $\overline \xi_3$ at $T=12T_0$ in {\bf case (\romannumeral1)}. (c) $\overline\xi_3$ versus photon energy $\varepsilon_\gamma$ obtained from {\bf case (\romannumeral1)} and {\bf case (\romannumeral4)}, respectively, where we switch off the pair production module while include spin and polarization effects in {\bf case (\romannumeral4)}.}
\end{figure*}

By comparing the positron number in {\bf case (\romannumeral2)} and {\bf case (\romannumeral3)} in figure~\ref{fig2}(a), one also notices that the positron yield is reduced by about $2.5\%$ due to the pure spin effect. This reduction is caused by the local spin polarization of electrons in the EMSW. When electrons just enter the magnetic region, their initial spins are mainly parallel to the direction of magnetic field (spin-up, defined by $\bm S_i \cdot \bm e_2>0$) [see figure~\ref{fig1}(c)], owing to the radiative spin polarization in the previous magnetic region. The spin-dependent photon emission probability is therefore reduced, especially for high-energy photons [see the third term $-uK_{\frac{1}{3}}(y)(\bm S_i \cdot \bm e_2)$ at the right hand side of equation~(\ref{eq1}) and figure~4(a) in Ref.~\cite{Song2019pra}]. Then, the electron spin gradually flips to be anti-parallel to the magnetic field direction (spin-down, defined by $\bm S_i \cdot \bm e_2<0$). After entering the next magnetic region, the direction of magnetic field $B_z$ is reversed. The electron spin direction becomes parallel to the $B_z$ direction again, and consequently the same process arises with slightly weaker photon emission. The relative deviation of emitted photon number $\delta_\gamma^{\rm\romannumeral3-\romannumeral2}$ is plotted in figure~\ref{fig2}(c). One can see the absolute value of $\delta_\gamma^{\rm\romannumeral3-\romannumeral2}$ increases with the increase of photon energy $\varepsilon_\gamma$ when $\varepsilon_\gamma<0.4$ GeV. These high-energy $\gamma$ photons more likely decay into $e^+e^-$ pairs, leading to a slightly smaller positron yield in figure~\ref{fig2}(a). The spin effect discussed above is five times weaker than the photon polarization effect in terms of positron yield, and therefore we will mainly focus on the latter one. However, we note that the polarization of high-energy emitted photons highly relies on the spin of emitting electrons, hence the local spin polarization could have a prominent impact on the high-energy positron yield [see the next subsection].


\subsection{Photon polarization properties}

The remarkable difference of the positron yield between {\bf case (\romannumeral1)} and {\bf case (\romannumeral2)} [see figures~\ref{fig2}(a) and \ref{fig2}(b)] mostly originates in the highly polarized $\gamma$ photons. Figures~\ref{fig3}(a) and \ref{fig3}(b) present spatial distributions of the photon density $n_\gamma$ and average photon polarization $\overline\xi_3$ at $t=12T_0$ in {\bf case (\romannumeral1)}, in which both spin and polarization effects are included. The other two polarization components $\overline\xi_1$ and $\overline\xi_2$ are nearly zero, so not shown here. $\overline\xi_3$ is rather uniform in space with a positive average value of about 0.58, indicating that photons are emitted predominantly with a linear polarization oriented along $\bm e_1$, i.e. always in the $x$-$y$ plane.

\begin{figure*}[t]
\centering
\includegraphics[width=0.8\textwidth]{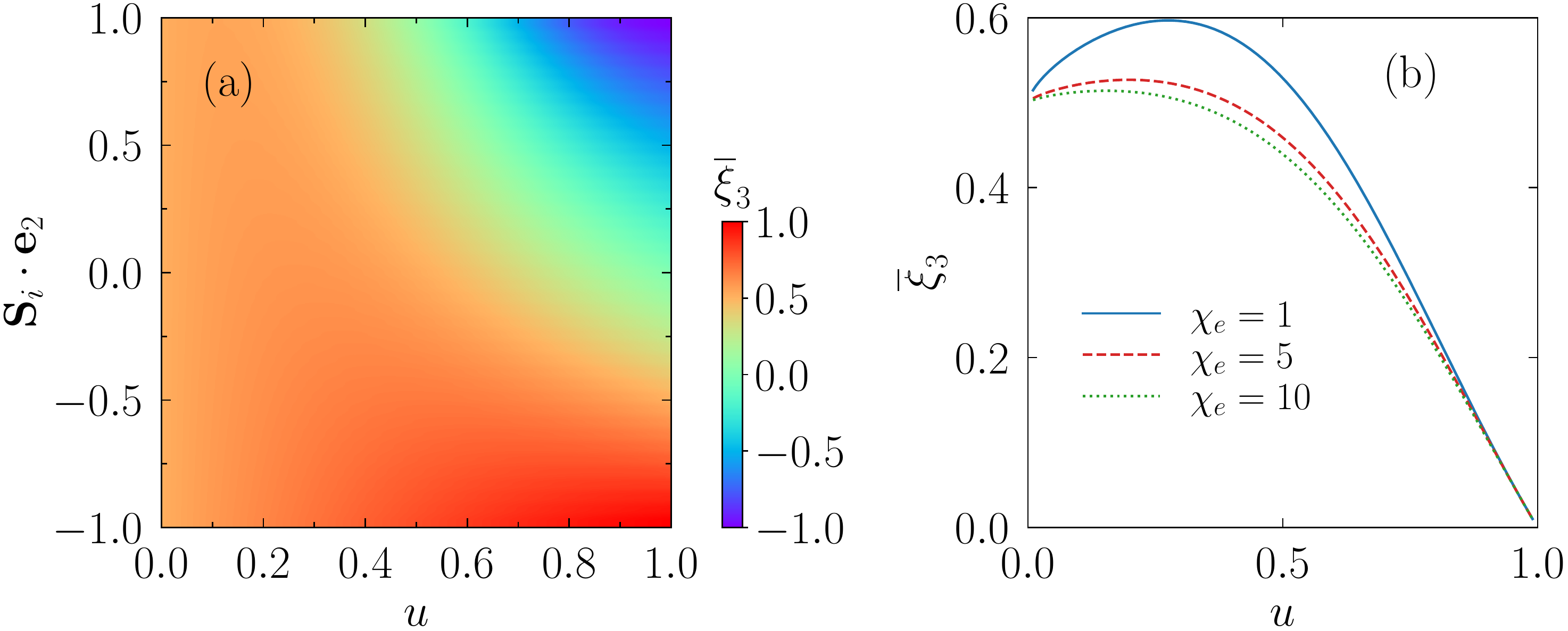}
\caption{\label{fig4} Theoretical calculations according to equation~(\ref{eq4}). (a) $\overline\xi_3$ versus emitted photon energy ratio $u$ and the initial electron spin vector $\bm S_i \cdot \bm e_2$ for $\chi_e=1$. (b) $\overline\xi_3$ versus $u$ at $\bm S_i \cdot \bm e_2=0$ for $\chi_e=$1, 5, and $10$.}
\end{figure*}

The spin- and polarization-resolved probabilities in equations~(\ref{eq1}) and (\ref{eq2}) can be simplified in the interaction between the considered linearly-polarized EMSW and un-prepolarized electrons. It is appropriate to sum up over $\bm S_f$ terms and neglect $\bm S_i \cdot \bm e_1$ and $\bm S_i \cdot \bm e_v$ terms in equation~(\ref{eq1}) since electrons can only be spin-polarized along $\bm e_2$, i.e. $B_z$ direction. Therefore, the average Stokes parameters of emitted photons can be approximately as
\begin{eqnarray}\label{eq4}
\overline\xi_1& \approx & 0,\nonumber\\
\overline\xi_2& \approx & 0,\\
\overline\xi_3& \approx & \frac{K_{\frac{2}{3}}(y) - \frac{u}{1-u} K_{\frac{1}{3}}(y)(\bm S_i \cdot \bm e_2)}{\frac{u^2-2u+2}{1-u}K_{\frac{2}{3}}(y)-{\rm Int}K_{\frac{1}{3}}(y) -uK_{\frac{1}{3}}(y)(\bm S_i \cdot \bm e_2)}.\nonumber
\end{eqnarray}
Equation~(\ref{eq4}) indicates that the emitted photons cannot be circularly polarized since $\overline\xi_2=0$. According to equation~(\ref{eq4}), the theoretical average polarization $\overline\xi_3$ as a function of $u$ and $\bm S_i \cdot \bm e_2$ is shown in figure~\ref{fig4}(a). For the low-energy photon, $\overline \xi_3$ is always positive with a value of about $0.5$, insensitive to the initial spin vector $\bm S_i$ of the emitting electron. While for the high-energy photon, $\overline\xi_3$ strongly depends on $\bm S_i \cdot \bm e_2$. Spin-up electrons are prone to emit photons linearly polarized along $\bm e_2$ axis ($z$ axis in our case) with $\overline\xi_3<0$. While for spin-down electrons, one can obtain an opposite result of $\overline\xi_3>0$. In general, $\overline\xi_3$ value decreases with the increase of photon energy $\varepsilon_\gamma$. This trend is supported by the curve for {\bf case (\romannumeral1)} shown in figure~\ref{fig3}(c), that $\overline\xi_3 \approx 0.5 $ at $\varepsilon_\gamma=0.1$ GeV and only $\overline\xi_3 \approx 0.1 $ at $\varepsilon_\gamma=0.6$ GeV.



The transformation of  Stokes parameters from $(\xi_1, \xi_2, \xi_3)$ to $(\xi_1', \xi_2', \xi_3')$ can also be simplified in our case with the linearly-polarized EMSW. We can obtain the transformation angle $\theta\approx0$ or $\pi$ for the fact that the acceleration direction and velocity direction of electrons are both well confined in the $x$-$y$ plane (laser polarization plane). Consequently, $\xi_3$ with respect to the emitted frame ($\bm e_1,\bm e_2,\bm e_v$) can be directly substituted into equation~(\ref{eq2}), i.e. $\xi_3\approx{\xi_3'}$. Hence, the positive values of $\xi_3$ directly leads to the reduction of the positron yield.

The photon polarization and positron yield at their higher energy range exhibit some anomalous phenomena compared with those at their lower energy range, due to the local spin polarization of electrons. As outlined above, the most intense photon emissions occur in the early stage of each magnetic region, in which the spin-up emitting electrons dominate. The resulting decrease in the photon yield has been observed in figure~\ref{fig2}(c). Besides the photon yield, the polarization of high-energy photons is also strongly affected by these locally spin-polarized electrons. Figure~\ref{fig3}(c) for {\bf case (\romannumeral4)} indicates the original emitted photons with $\varepsilon_\gamma>0.5$ GeV possess the negative $\overline\xi_3$. It is coincident with the theoretical analysis shown in figure~\ref{fig4}(a) that spin-up electrons are in favor of emitting high-energy photons of negative-value $\overline\xi_3$. This eventually leads to an increase of positron yield at positron energies higher than 0.5 GeV, rather than decreasing like that at lower energies in figure~\ref{fig2}(b), due to the opposite signs of $\overline\xi_3$ between these two energy ranges. Moreover, when we switch on the pair production process and include the $\gamma$-photon annihilation in {\bf case (\romannumeral1)}, the polarization $\overline\xi_3$ of high-energy photons increases as compared to those in {\bf case (\romannumeral4)}, shown in figure~\ref{fig3}(c). This is because photons of negative $\xi_3$ are easier to annihilate into $e^-e^+$ pairs, according to the polarization-dependent pair production probability in equation~(\ref{eq2}).

\begin{figure*}[t]
\centering
\includegraphics[width=0.5\textwidth]{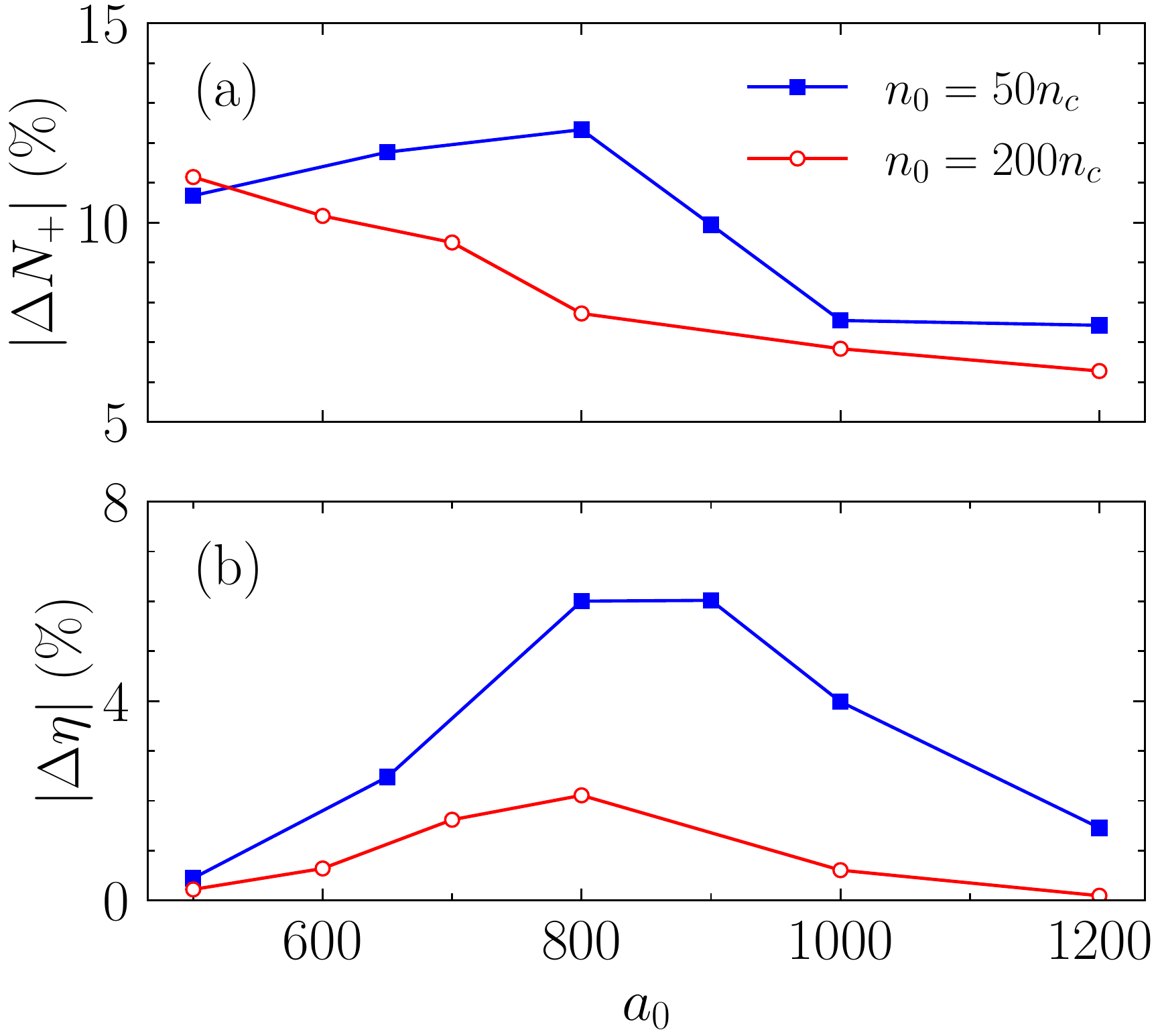}
\caption{\label{fig5} The absolute relative deviation of (a) positron number $|\Delta N_{+}|=|(N_+^{\rm inc}-N_+^{\rm exc})/N_+^{\rm exc}|$ and (b) total laser absorption $|\Delta\eta|=|(\eta^{\rm inc}-\eta^{\rm exc})/\eta^{\rm exc}|$ between including and excluding spin and polarization effects, under various laser strengths $a_0=500$-$1200$ and initial plasma densities $n_0/n_c=50$ and $200$.}
\end{figure*}

\subsection{Impacts of laser intensity and plasma density}

In figure~\ref{fig5}(a), we investigate the difference of positron yield between with and without spin and polarization effects under various laser strengths $a_0$ in order to find the laser intensity at which these two effects need to be taken into account. For the case of initial plasma density $n_0=50n_c$, the absolute relative deviation $|\Delta N_+|$ of total positron number first increases with laser intensity $a_0$ and reaches a maximum $12\%$ at $a_0=800$, then it gradually decreases to $7\%$ at $a_0=1200$. For a higher-density plasma of $n_0=200n_c$, $|\Delta N_+|$ always decreases with $a_0$ in the scanned parameter range, i.e. decreasing from $11\%$ at $a_0=500$ to $6\%$ at $a_0=1200$. On the whole, $|\Delta N_+|$ is larger in the case of $n_0=50n_c$ than that of $n_0=200n_c$ case at the same $a_0$. These results can be explained according to figure~\ref{fig4}(b), in which the average polarization $\overline\xi_3$ as a function of $u$ under different quantum parameter $\chi_e$ is plotted according to equation~(\ref{eq4}). Here, we assume  $\bm S_i\cdot\bm e_2=0$, which is approximately valid since electrons or positrons cannot gain a net degree of spin polarization in the linearly EMSW and the influence of local spin polarization on the positron yield is smaller compared with the photon polarization one. One can see that the average polarization $\overline\xi_3$ is smaller for a higher $\chi_e$ (corresponding to a higher laser intensity), hence a smaller difference between positron yields can expected.

Figure~\ref{fig5}(b) shows the spin and polarization effects on the total laser absorption $\eta$ with laser strength $a_0$. The absolute relative deviation $|\Delta\eta|$ can reach a maximum of $6\%$ for $n_0=50n_c$ and $2\%$ for $n_0=200n_c$, respectively. Below $a_0=800$, $|\Delta\eta|$ decreases as reducing $a_0$ and only $\Delta\eta<0.5\%$ is observed at $a_0=500$, indicating that the spin and polarization effects have a negligible impact on the laser absorption at laser intensity below $I_{\rm min}=3.5\times 10^{23}$ W/cm$^2$, although the difference of positron yield is more obvious at relatively low laser intensities. This is because the pair production probability $d^2W_{pairs}/({d\varepsilon_{+} dt})$ in  equation~(\ref{eq2}) is exponentially small for $\chi_\gamma\ll 1$, so that only a small number of positrons is produced. At $I_0<I_{\rm min}$, the energy conversion efficiency from laser to positron is less than $1\%$, hence the impact of positron yield reduction induced by spin and polarization effects on the laser-plasma interaction can be neglected.


\section{Conclusion}
\label{conclusion}
In conclusion, we have investigated the pair production in the interaction of two counter-propagating 10-PW-class laser pulses with a thin foil target using QED-PIC simulations, with $e^-e^+$ spin and $\gamma$-photon polarization effects included. The two effects result in the decrease of total positron yield by about $10\%$, and the relative difference can even reach $20\%$ for the medium-energy positrons. In other words, the spin- and polarization-averaged probabilities widely implemented in QED-PIC methods overestimate the positron yield, which can also affect the laser plasma interaction at laser intensities above $3.5\times 10^{23}$ W/cm$^2$ in our laser-foil interaction. The decrease of positron yield mainly comes from the linear polarization up to $50\%$ of emitted $\gamma$ photons. In addition, we also observe several anomalous phenomena about positron yield and photon polarization of high-energy particles caused by the local spin polarization of $e^-e^+$. For example, with spin and polarization effects included, the positron yield is increased in the high-energy range rather than decreased like that in the low-energy range.

\begin{figure*}
\centering
\includegraphics[width=0.7\textwidth]{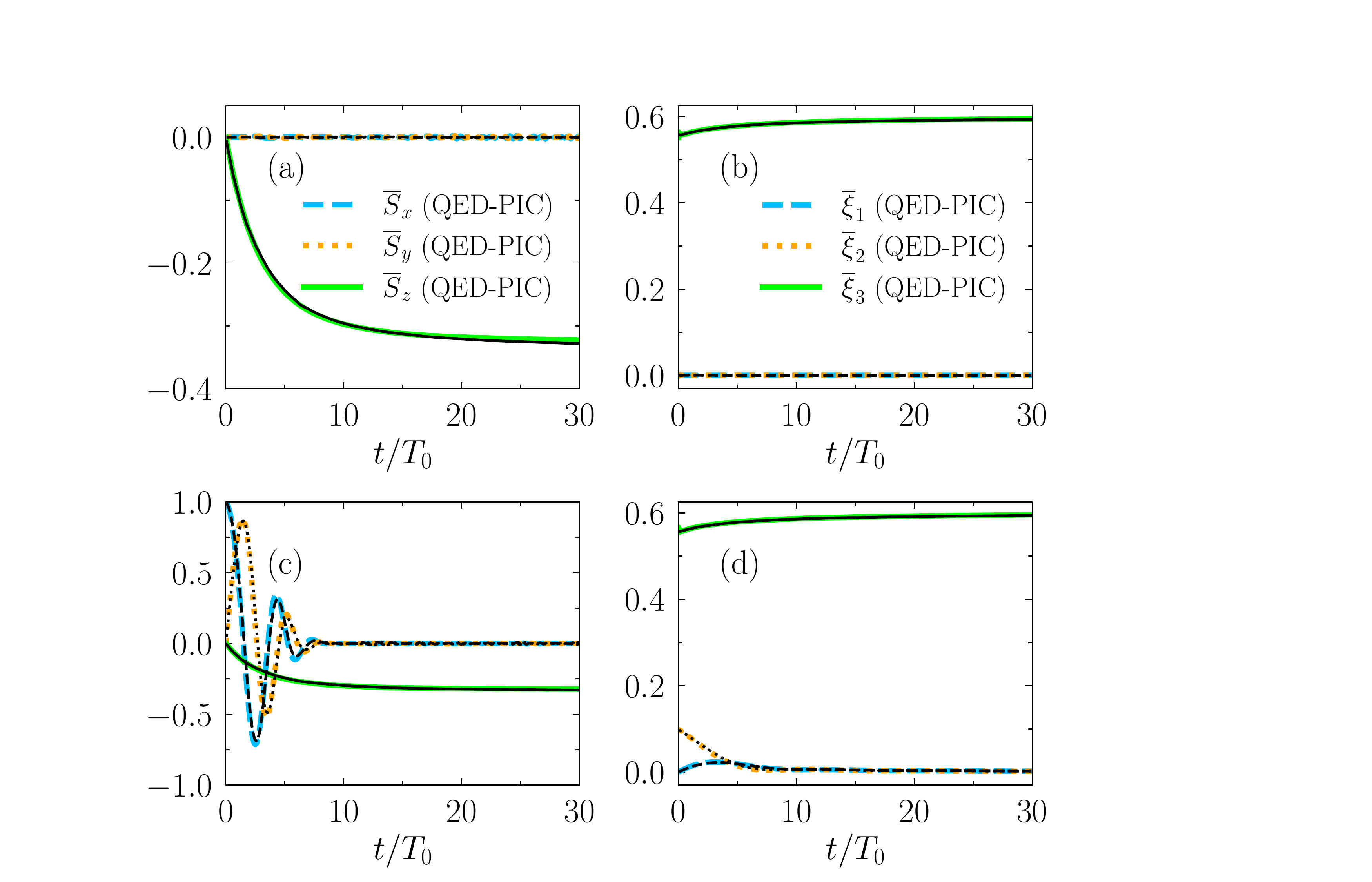}
\caption{\label{fig6} Comparison of our 2D QED-PIC results (thick colored lines) with those by the code in reference~\cite{Li2020prl2} (thin black lines) in the process of radiative spin polarization of initially (a)(b) unpolarized electrons and (c)(d) longitudinally polarized electrons, respectively. (a)(c) Average spin degree of electrons $\overline S_x$, $\overline S_y$ and $\overline S_z$. (b)(d) Average polarization degree of all emitted photons $\overline \xi_1$, $\overline \xi_2$ and $\overline \xi_3$ with respect to the emitted frame.}
\end{figure*}

\section*{Acknowledgments}
\addcontentsline{toc}{section}{Acknowledgments}

This work was supported by the National Key R\&D Program of China
(Grant No. 2018YFA0404801), National Natural Science Foundation of
China (Grant Nos. 11775302, 11721091, 11991073, and 12075187), the Strategic Priority
Research Program of Chinese Academy of Sciences (Grant Nos.
XDA25050300, XDA25010300, and XDB16010200), Science Challenge
Project of China (Grant Nos. TZ2016005 and TZ2018005), and the
Fundamental Research Funds for the Central Universities, the
Research Funds of Renmin University of China (20XNLG01).

\appendix

\section{Spin and polarization benchmarks}
\label{appendix}

In our QED-PIC code, the $e^-e^+$ radiative spin polarization and photon polarization determination modules follow the Refs.~\cite{Cain,Li2020prl1,Li2020prl2}, in which the three-dimensional spin dynamics can be resolved. Here, we perform two additional 2D QED-PIC simulations to benchmark against the one-particle code used in reference~\cite{Li2020prl2}, where ultrarelativistic electrons with 1 GeV move along $+x$ axis in the $x$-$y$ plane under a perpendicularly static external magnetic field of $B_0=100mc\omega_0/|e|$, where $\omega_0=1\mu$m. We take the computational domain of $8\lambda_0\times8\lambda_0$ in $x\times y$ directions with $128\times 128$ cells, and 16 macro electrons per cell. The electron density is low enough to avoid the influence of self-generated electromagnetic field, and periodic boundaries are employed. Figures~\ref{fig6}(a) and \ref{fig6}(b) show the time evolution of average spin degree  $\bm{\overline S}$ of electrons and average polarization degree $\bm{\overline \xi}$ of emitted photons for the initially unpolarized (or randomly polarized) electrons, respectively, and figures~\ref{fig6}(c) and \ref{fig6}(d) are those for initially longitudinally polarized electrons. Our QED-PIC simulation results are in good agreement with the single-particle model simulation results of the code in reference~\cite{Li2020prl2}.

\end{document}